\documentclass[12pt]{article}
\usepackage{comment}
\usepackage{verbatim}
\usepackage{times}
\usepackage{epsfig,amsmath}
\usepackage{subfigure}
\usepackage{graphicx}
\usepackage{color}
\usepackage{epstopdf}
\usepackage[dvipsnames]{xcolor}
\usepackage[colorlinks, linkcolor=OrangeRed,citecolor=RoyalBlue,urlcolor=NavyBlue]{hyperref}
\usepackage[all]{hypcap} 

\usepackage{microtype,physics,cleveref,amsmath,amssymb,hyphenat}
\usepackage[backend=biber,sorting=none,style=numeric-comp]{biblatex}
\newenvironment{sciabstract}{%
\begin{quote} \bf}
{\end{quote}}

\newcounter{lastnote}

\title{Non-Hermitian Exceptional Topology on a Klein Bottle Photonic Circuit}

\author
{Ze-Sheng Xu,$^{1,\ast}$ J. Lukas K. König,$^{2,\ast}$ Andrea Cataldo,$^{1}$ Rohan Yadgirkar$^{1}$, \\ Govind Krishna,$^{1}$ Venkatesh Deenadayalan$^{3}$, Val Zwiller$^{1}$,\\ Stefan Preble$^{3}$, Emil J. Bergholtz$^{2}$, Jun Gao$^{1,4,\ast}$, \\ Ali W. Elshaari$^{1,\ast}$\\
\\
\normalsize{$^1$Department of Applied Physics, KTH Royal Institute of Technology, Albanova University Centre,}\\
\normalsize{Roslagstullsbacken 21, 106 91 Stockholm, Sweden}\\
\normalsize{$^2$Department of Physics, Stockholm University, AlbaNova University Center, } \\
\normalsize{106 91 Stockholm, Sweden}\\
\normalsize{$^3$Microsystems Engineering, Rochester Institute of Technology, } \\
\normalsize{1 Lomb Memorial Drive, Rochester, NY 14623}\\
\normalsize{$^4$School of Optical and Electronic Information, Huazhong University of Science and Technology,}\\ 
\normalsize{Luoyu Road 1037, Wuhan, Hubei, 430074, China}\\
\normalsize{$^\ast$E-mail: zesheng@kth.se, lukas.konig@fysik.su.se, jungao@hust.edu.cn, elshaari@kth.se}
}

\date{}

\begin{document}
\baselineskip24pt

\maketitle

\section*{Abstract}
Non-Hermitian physics has unlocked a wealth of unconventional wave phenomena beyond the reach of Hermitian systems, with exceptional points (EPs) driving enhanced sensitivity, nonreciprocal transport, and topological behavior unique to non-Hermitian degeneracies. Here, we present a scalable and reconfigurable silicon photonic integrated circuit capable of emulating arbitrary non-Hermitian time evolution with high precision. Using this programmable platform, we implement a two-band non-Hermitian Hamiltonian defined on a Klein-bottle topology—a nonorientable parameter space that enables exceptional phases forbidden on orientable manifolds. Through an on-chip amplitude-and-phase reconstruction protocol, we retrieve the full complex Hamiltonian at multiple points in parameter space and experimentally map the associated Fermi arc where the imaginary eigenvalue gap closes. The orientation of the measured Fermi arc reveals a nontrivial exceptional topology: it implies the presence of same-charge EPs (or an EP monopole) that cannot annihilate locally on the Klein bottle. Our results demonstrate the first photonic realization of exceptional topology on a nonorientable manifold and establish a versatile platform for exploring exotic non-Hermitian and topological models relevant to classical and quantum photonics.

\begin{sciabstract}
\end{sciabstract}

\section*{Introduction}
Non-Hermitian physics, originally motivated by the theoretical modeling of open systems with intrinsic gain and loss\cite{ashida2020non, el2018non, kawabata2019symmetry, ruter2010observation}, has emerged as a powerful framework for exploring unconventional wave phenomena beyond the scope of traditional Hermitian systems. In recent years, it has demonstrated striking capabilities such as non-Hermitian enhanced sensitivity\cite{chen2017exceptional, wiersig2014enhancing, wiersig2020review, de2022non, hodaei2017enhanced, zhong2019sensing, lai2019observation, budich2020non}, nonreciprocal transport\cite{lin2011unidirectional,mandal2020nonreciprocal, peng2014parity, chang2014parity, feng2011nonreciprocal}, and the realization of non-Hermitian-assisted lasing\cite{siegman1989excess,wong2016lasing, hodaei2014parity, feng2014single, teimourpour2016non, hokmabadi2019supersymmetric}. Non-Hermitian degeneracies, so-called exceptional points (EPs), play a particularly pivotal role among these various features. An EP is characterized by a simultaneous coalescence of both eigenvalues and their corresponding eigenstates, leading to a reduction in the dimensionality of the eigenspace. This is in stark contrast to diabolic points (DPs) in Hermitian systems, where eigenvalues become degenerate while the associated eigenstates remain orthogonal\cite{heiss2004exceptional,bergholtz2021exceptional, ding2022non}. The non-diagonalizable nature of EPs originates from a topological classification in terms of eigenvalue homotopy 
\cite{wojcikHomotopyCharacterizationNonHermitian2020,liHomotopicalCharacterizationNonHermitian2021,yangHomotopySymmetryNonHermitian2024} and results in nontrivial topological structures in parameter space, such as branch point singularities and associated Riemann surfaces, which underlie a range of exotic physical effects.

Implementing non-Hermitian systems, and their accompanying non-unitary dynamics, requires a new suite of platforms.
With the rapid advancements in photonics, micro- and nano-fabrication techniques, as well as semiconductor technology, reconfigurable PICs have emerged as a versatile and scalable platform for implementing arbitrary linear optical transformations with high precision and long-term stability\cite{bogaerts2020programmable, gyger2021reconfigurable}. By leveraging networks of Mach-Zehnder interferometers (MZIs) with tunable phase shifters\cite{bogaerts2018silicon}, these systems can realize any unitary operation within a given mode space\cite{reck1994experimental,miller2013self, carolan2015universal,clements2016optimal}. This makes them suitable for a wide range of applications, including telecommunication, optical ranging, quantum information processing, and fundamental wave physics -- particularly in the study of topological photonics\cite{dai2024programmable, on2024programmable}. The programmability of these circuits allows for in situ reconfiguration of optical pathways, enabling rapid prototyping and systematic exploration of diverse photonic Hamiltonians. However, conventional implementations of non-Hermitian photonic systems typically rely on the use of nonlinear optical materials to realize gain and loss. This poses a significant challenge for simulating non-Hermitian models within linear, Complementary Metal-Oxide-Semiconductor (CMOS)-compatible silicon PICs. As a result, realizing non-Hermitian physics -- particularly the tunable and programmable kind required for exploring complex topological features -- in such a linear platform demands alternative strategies beyond direct optical amplification or attenuation.

Instead of introducing actual optical gain to realize the non-Hermitian model, we employ an ancilla-mode approach. In this work, we implement a differential-loss scheme to simulate the time evolution under non-Hermitian Hamiltonians. This approach is mathematically equivalent to applying a global decay factor to the evolution operator derived from the gain/loss Hamiltonian, with the decay set by a tunable imaginary diagonal term in the effective Hamiltonian \cite{ornigotti2014quasi}. By introducing ancilla modes at the periphery of the system, the desired non-unitary transformation is embedded into a higher-dimensional unitary, allowing non-Hermitian dynamics to be emulated within a fundamentally linear photonic architecture \cite{tischler2018quantum}. Within this framework, we construct a programmable photonic circuit capable of realizing arbitrary \(2\times2\) non-Hermitian Hamiltonian dynamics. Using this silicon PIC, we experimentally demonstrate time evolution governed by non-Hermitian models featuring EPs or merged EPs with monopole structure in a two-dimensional nonorientable momentum space, exemplified by the Klein bottle. We further show how essential information about the underlying Hamiltonian can be retrieved from measured output intensity profiles. Our work demonstrates the feasibility of exploring non-Hermitian physics within linear PICs and offers pathways toward on-chip emulation of diverse physical models.

\section*{Model}

We observe a two-dimensional gapless non-Hermitian phase that breaks the fermion doubling theorem due to an underlying non-orientable parameter space \cite{konig2025exceptional}, which in particular hosts a pair of EPs with the same topological charge that cannot annihilate locally.

The parameter space we consider, the Klein bottle, can be thought of as the fundamental domain of a momentum-space Hamiltonian \(H(p,q)\) subject to a non-symmorphic symmetry that identifies \(H(p,q)=H(-p,q+\pi)\). Such symmetries can arise in momentum space through, for example, real-space gauge structure \cite{chenBrillouinKleinBottle2022, liKleinbottleQuadrupoleInsulators2023,puAcousticKleinBottle2023,wangChessboardAcousticCrystals2023,taoHigherorderKleinBottle2024, zhuBrillouinKleinSpace2024,huHigherOrderTopologicalInsulators2024,huHigherOrderTopologicalInsulators2024,huAcousticHigherorderTopological2024,nguyenFermiArcReconstruction2023,laiRealprojectiveplaneHybridorderTopological2024,shenObservationKleinBottle2024,wangOpticalInterfaceStates2017,qiuOctupoleTopologicalInsulating2024,xiaoSpinSpaceGroups2024,calugaruMoireMaterialsBased2025}. 
On the Klein bottle, the Fermion doubling theorem holds only in a modified form, requiring the sum of all EP topological charges to be
\begin{equation}
    \sum_i B_{\text{EP}_i} = B_p B_q B_p B_q^{-1},
\end{equation}
a specific combination of two topological charges \(B_{p,q}\) that are associated to the parameter space globally, rather than any individual EP.
These topological charges are elements in the braid group. For the case of two bands considered here, these simplify to a winding number invariant \(W\in\mathbb{Z}\). 
The sum rule thus turns into a parity requirement
\begin{equation}
    \sum_i W_{\text{EP}_i} = 2 W_p.
\end{equation}
A phase with non-zero invariant, say \(W_p=1\), can therefore not be gapped. 
Instead it generically hosts two EPs, with topological charges \(W_{\text{EP}_{1,2}}=1\). 
By tuning the system, these can be brought to fuse, but rather than annihilating, they will produce a monopole with charge \(W_\text{MP}=2\).
Such arrangements of degeneracies are strictly forbidden in Hermitian systems, and even in non-Hermitian systems on orientable backgrounds.

We realize the Hamiltonian 
\begin{equation}\label{eq4}
H_{\mathrm{EP}}^{\mathrm{\mathbf{K}}^2}(p,q)=
\mqty(
-a_{p,q} & c \\
c & a_{p,q}
)
\end{equation}
with $a_{p,q} = \sin p \cos q + i(1-\cos p)(1+\cos(2q)/2)$ a detuning defined on the \((p, q)\) parameter space plane, and $c$ the coupling strength between the two modes. The Hamiltonian is subject to the non-symmorphic symmetry \(H_{\mathrm{EP}}^{\mathrm{\mathbf{K}}^2}(p,q)=H_{\mathrm{EP}}^{\mathrm{\mathbf{K}}^2}(-p,\pi+q)\), which reduces its fundamental domain to $(p,q) \in [-\pi,\pi]\times [0,\pi]$, forming a Klein bottle due to the non-orientable boundary identification \((p,0)=(-p,\pi)\). The topological phase invariant \(W_p\) is determined by the eigenvalue winding along \(p\) as \(q=0\). In this experiment, we investigate the light evolution for different values of $c$ ranging from 1 to 12, demonstrating high fidelity, and further explore the monopole topology with a particular focus on $c$ = 1. The robust nature of EPs is also encoded in the topological bulk Fermi arcs, the lines in parameter space at which the imaginary part of the eigenvalue gap closes \cite{li2011origin, zhou2018observation, schrunk2022emergence}.
These terminate at the EPs, and reverse their orientation when crossing the \(q=0\)-boundary, enabling a single, consistently oriented Fermi arc to connect two EPs with winding \(W_{\text{EP}_{1,2}}=1\).

Given a non-Hermitian Hamiltonian \(H\), we simulate the evolution under the pure-loss Hamiltonian \(H'=H-i \Lambda \mathbb{I}\), for \(\Lambda>0\) chosen sufficiently large such that all eigenvalues of \(H\) lie in the lower-half complex plane. 
This transformation does not affect the eigenstates; the dynamics of the original \(H\) can be obtained from the measured results by multiplying all measured intensities at evolution time \(t\) by a factor \(e^{\Lambda t}\). Throughout the remaining manuscript, the parameter $\Lambda$ will be set to $3$ to derive the differential-loss Hamiltonian from $H_{\mathrm{EP}}^{\mathrm{\mathbf{K}^2}}$, ensuring that this resulting Hamiltonian is physically realizable using only loss mechanisms.

\section*{Methods}

\begin{figure}[htbp]
	\centering
	\includegraphics[width=0.8\linewidth]{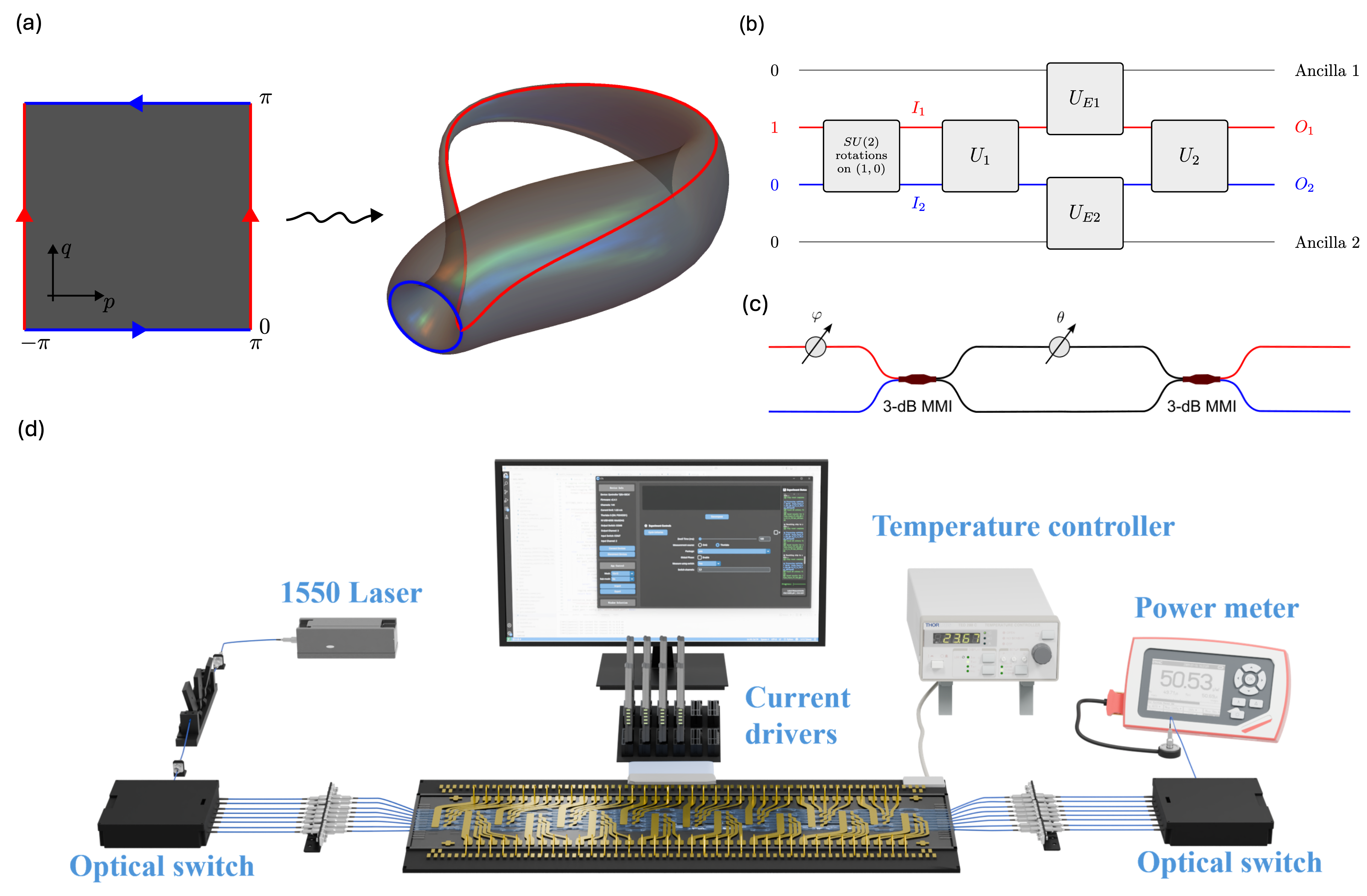}
	\caption{\textbf{Realization of non-Hermitian model and topology.} 
    \textbf{(a)}~\textit{Exceptional phases on nonorientable manifolds.} Under the shown boundary identifications, the free parameter space rectangle transforms into a Klein bottle, the nonorientable momentum manifold on which the non-Hermitian Hamiltonian is defined. 
    \textbf{(b)}~The matrices \( U_1 \), \( U_2 \), \( U_{E1} \), and \( U_{E2} \) form a unitary \( 4 \times 4 \) matrix that simulates the dynamics of a non-Hermitian \( 2 \times 2 \) evolution operator  \( T \) , with coupling to the environment implemented through ancilla modes~1 and~2. The components \( U_1 \) and \( U_2 \) correspond to tunable beam splitters and phase shifters acting within the system, while \( U_{E1} \) and \( U_{E2} \) couple the system to the environment, represented here by the ancilla modes.
    \textbf{(c)}~Schematic of a single Mach-Zehnder interferometer (MZI) unit, serving as the fundamental computational element of the programmable photonic circuit. The brown boxes represent 3-dB MMIs, and the circles with arrows denote the external and internal phase shifters controlling $\phi$ and $\theta$, respectively.\textbf{(d)}~An illustration of the experimental setup is shown, where the sizes of individual components have been adjusted for clarity. A 1550 nm pulsed diode laser (PDL 800-B) is employed as the light source. The polarization of the transmitted light is controlled to be in the TE mode using a three-paddle polarization controller. Two $1\times12$ MEMS optical switches are used in the system (Use eight of the channels). The input light is coupled into the first optical switch, allowing any of the eight operational modes or four additional test modes to be selected under computer control.The core of the system is an $8\times8$ silicon PIC comprising an array of 28 Mach-Zehnder Interferometers (MZIs) and 56 heating-electrodes. The PIC is controlled by one of eight current driver modules (Q8iv, Qontrol Ltd.) interfaced with the host computer running the GUI based on $Python$ 3.11. The output light is coupled into the second optical switch and its intensity is measured using a Thorlabs PM400 power meter. During the measurement process, the second optical switch rapidly cycles through output modes 1 to 8 at each experimental step to acquire the full set of output intensities. }
	\label{f1}
\end{figure}

To simulate the non-unitary evolution of systems governed by effective non-Hermitian Hamiltonians, we employ a dilation scheme based on the singular value decomposition (SVD) \cite{eckart1936approximation}.
We achieve this using a circuit structure composed of four MZIs arranged as shown in \Cref{f1}~(b);
two unitary transformations ($U_1$ and $U_2$) are interspersed with the auxiliary MZIs ($U_{E1}$ and $U_{E2}$), which couple the system to an ancilla that is subsequently discarded. The evolution is measured in two nominal modes \cite{tischler2018quantum}.
We refer the reader to the \textbf{Supplemental Information} for a more detailed derivation.  The minimum computational unit in this PIC is a Mach-Zehnder interferometer (MZI) [see \Cref{f1}~(c)], consisting of two multi-mode interferometers (MMI) and bending waveguides. 
Each MZI implements the $2\times2$ unitary scattering matrix
\begin{equation}\label{eq3}
    U_\text{MZI}(\theta,\phi) =  e^{i\frac{\theta+\pi}{2}}
        \mqty[
     e^{i\phi}\sin(\frac{\theta}{2})&  \cos(\frac{\theta}{2})\\
     e^{i\phi}\cos(\frac{\theta}{2})& -\sin(\frac{\theta}{2})
     ],
\end{equation}
realizing subsequent rotations in the Bloch sphere about the \(Z\) and \(Y\)-axes by angles \(\phi\) and \(\theta\), respectively determining the beam-splitting ratio and the relative phase between the two arms.
The angles $\theta$ and $\phi$ are independently controlled by an internal and external thermal phase shifter via electronic driver modules (Qontrol) interfaced with a host computer.
The MZI configurations $\theta=0$ and $\theta=\pi$ are defined as cross or bar states, respectively \cite{alexiev2021calibrating}. In our setup, these MZIs are arranged in an $8\times8$ programmable PIC as shown in \Cref{f1}~(d), fabricated via a CMOS-compatible silicon photonics process (SiPhotonIC ApS).
The device integrates 28 thermally tunable MZIs, which together form a reconfigurable mesh capable of implementing arbitrary linear transformations.

Modulation of these phase shifters enables fine-grained control ($50~\mu\text{A}$ current and $0.0125\pi$ steps) of the optical interference and power distribution across the mesh. We use a pulsed diode laser (PDL 800-B) emitting at 1550 nm as the coherent light source. To ensure excitation of the transverse electric (TE) mode within the PIC waveguides, we control its polarization state using a three-paddle fiber polarization controller. We route the laser output to the PIC through a \(1\times12\) Microelectromechanical Systems (MEMS) optical switch, enabling programmable selection from the principal input/output modes. Light exiting the PIC is routed through a second MEMS \(1\times12\) optical switch to a Thorlabs PM400 power meter for intensity readout. During each measurement cycle, the output switch rapidly scans across the eight output ports, allowing quasi-synchronous acquisition of the full output intensity distribution corresponding to a given input and PIC configuration. All optical interconnects are established via polarization-maintaining fibers, which couple light to/from the chip via V-groove fiber arrays. The electrical control and data transmission are handled through a hub-connected system. For optical microscope images of the device and the packaged circuit, we refer the reader to the \textbf{Supplemental Information}. With the described setup, we are able to realize arbitrary equivalent non-unitary dynamics. We sample the time evolution of a given Hamiltonian \(H\) by implementing a discrete series of evolution operators \(T_{t_n} = e^{-i H'  t_n}\) for a set of time steps \(\{t_i\}\). The phase evolution of the target channel can be measured through on-chip interferometric detection to extract the encoded information similar to the method described in \cite{zhang2021optical}. In this experiment, we designed a complete sequential measurement architecture to capture both intensity and phase information, with the detailed scheme provided in the \textbf{Supplemental Information}.

\section*{Results}

\begin{figure}[htbp]
	\centering
	\includegraphics[width=1.0\linewidth]{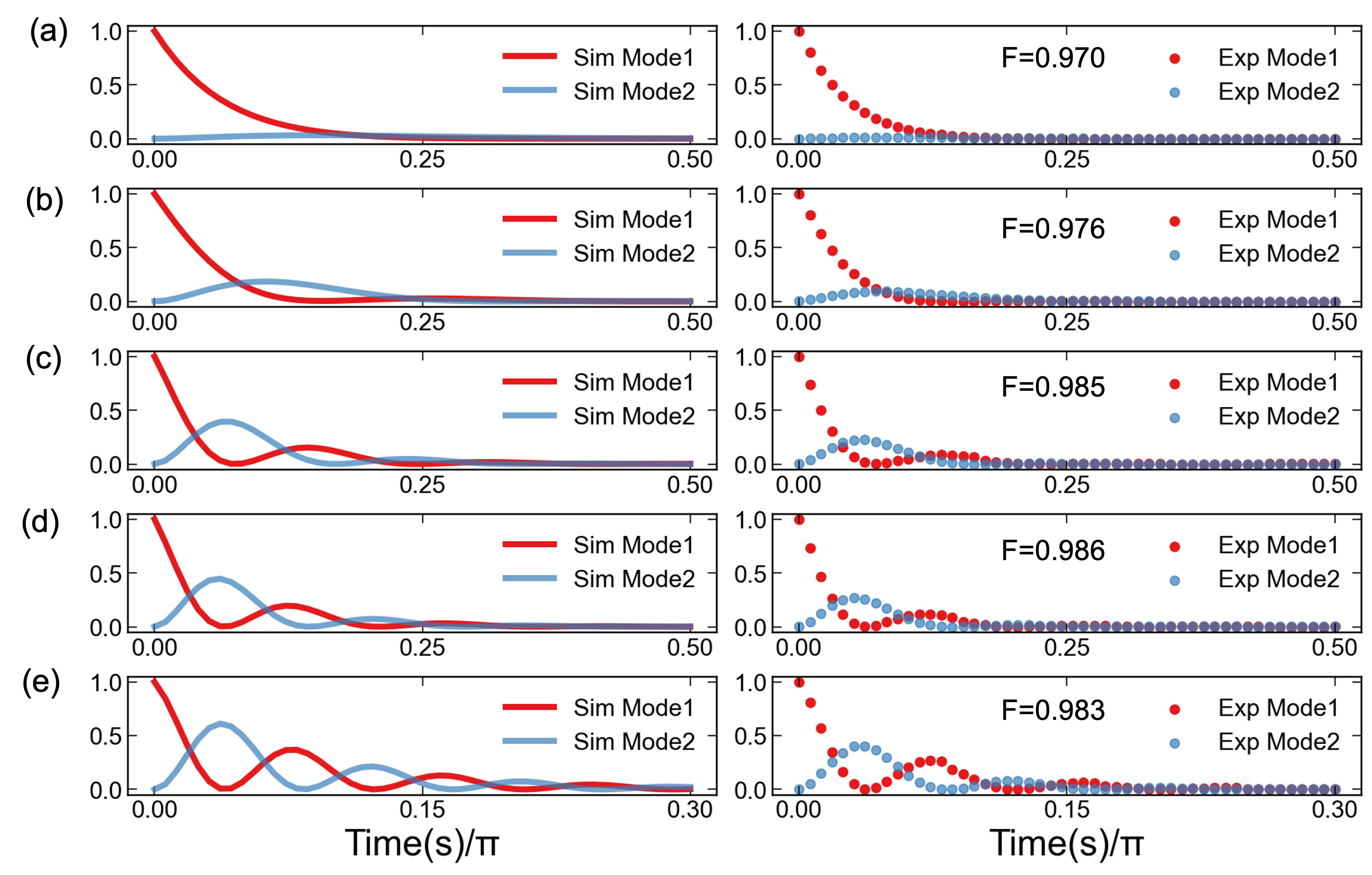}
	\caption{\textbf{Simulation (left) vs experiment (right) at five parameter points.}
	Each subpanel shows normalized intensities of mode~1 (red) and mode~2 (blue) versus normalized time; left column: numerical simulation, right column: experimental data.
	\textbf{Parameters}:
	\textbf{(a)} \(c=1\), \(p=0.5\pi\), \(q=0.51\pi\);
	\textbf{(b)} \(c=3\), \(p=0.2\pi\), \(q=0.986\pi\);
	\textbf{(c)} \(c=6\), \(p=\pi\), \(q=0.5\pi\);
	\textbf{(d)} \(c=7\), \(p=\pi\), \(q=0.5\pi\);
	\textbf{(e)} \(c=12\), \(p=\pi\), \(q=0.5\pi\).
	The Fidelity is shown as an inset. The measured traces closely match the simulations in oscillation and decay, confirming faithful implementation of the non-Hermitian dynamics at each \((c,p,q)\) point.}
	\label{f2}
\end{figure}

To verify the compatibility of our experimental setup with our models, we compared numerical simulations and experimental measurements of the output intensities of mode 1 and mode 2, as shown in \Cref{f2}. The data points (a)-(e) correspond to samples obtained under different combinations of the three parameters $[c, p, q] $ defined in \Cref{eq4}. 
To quantify this agreement, we evaluated the fidelity between the experimental data and the theoretical curves by computing weighted correlation coefficients, the details of which are provided in the \textbf{Supplemental Information} in addition to the calibration procedure of our setup.
We find that the correlation values exceed 0.97 for all selected parameter points, confirming the high fidelity of our experimental implementation.

The Fermi arc refers to the locus in parameter space where the difference between the eigenvalues vanishes either in real or imaginary part. To reconstruct the Fermi arc in our model and verify the characteristic properties of the non-Hermitian Hamiltonian, we employ the following measurement procedure: We first apply unitary dilation to embed the $2 \times 2$ non-unitary operator -- generated by the time evolution under the non-Hermitian Hamiltonian -- into $4 \times 4$, then   $6 \times 6$ unitary operators, which are then implemented on the PIC. The detailed procedure is provided in the \textbf{Supplemental Information}. The central idea is to fit the amplitudes and phases of the four matrix elements of the time-evolution $2 \times 2$ operator at several sampled time points, and from this fitting procedure extract the corresponding non-Hermitian Hamiltonian. Once the Hamiltonian is reconstructed, we compute the difference between its eigenvalues and thereby determine whether a given point lies on the Fermi arc.

\begin{figure}[htbp]
	\centering
	\includegraphics[width=1\linewidth]{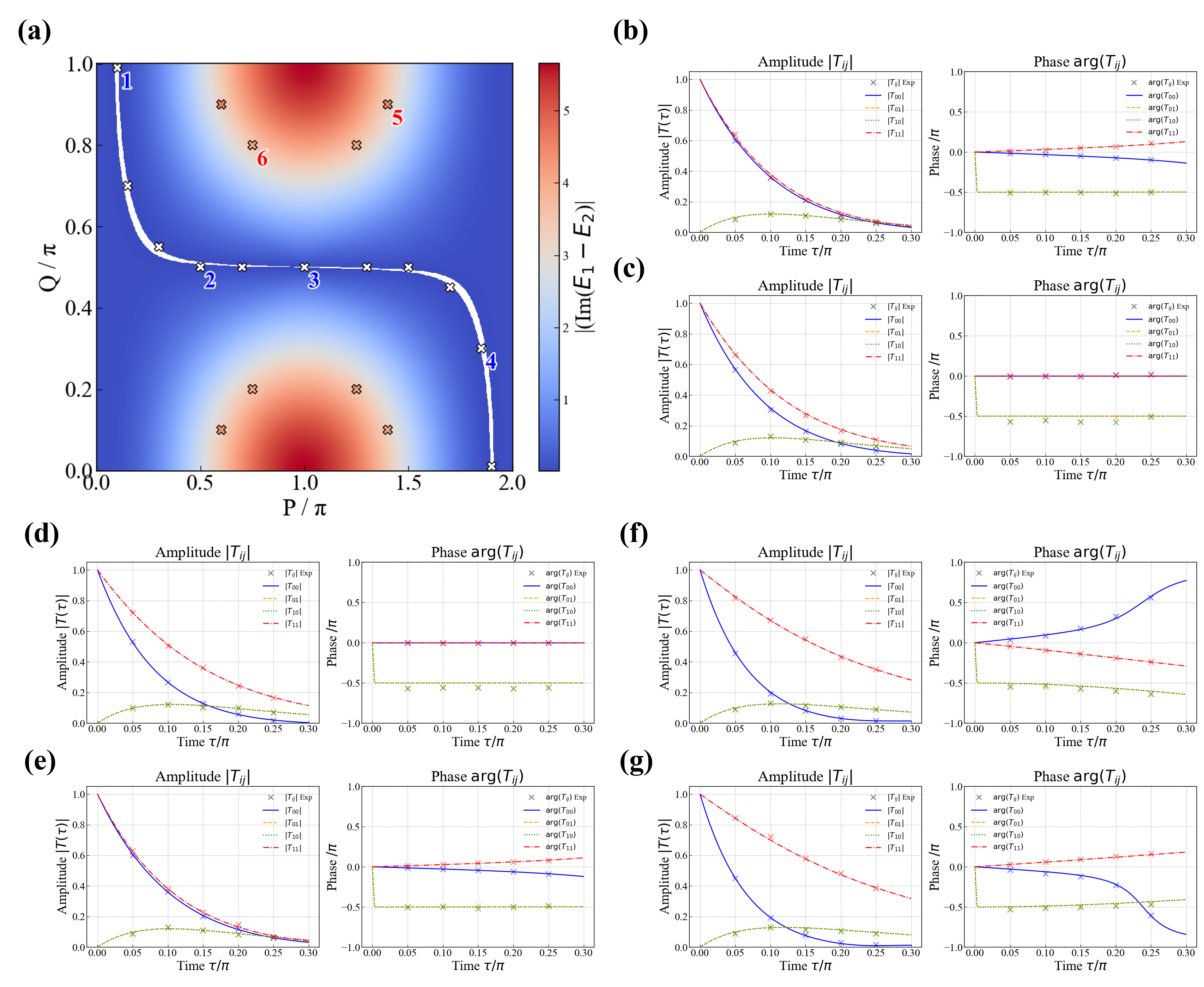}
	\caption{\textbf{Simulated and experimentally sampled Fermi arc with phase- and amplitude-resolved measurements.}
\textbf{(a)} Numerically computed bifurcation of the imaginary parts of the eigenvalues over the $(p,q)$ parameter space for $c=1$, together with the experimentally sampled parameter points. White markers indicate points that lie on the Fermi arc (imaginary-gap closure), while red markers denote points in gapped regions.  
\textbf{(b–g)} For each labeled parameter point, we measure the amplitudes and phases of all four elements of the time-evolution operator $T_{t_n}=e^{-iH' t_n}$ at five evolution times $t_n=\{0.05\pi, 0.10\pi, 0.15\pi, 0.20\pi, 0.25\pi\}$. Each subpanel shows the experimentally measured data together with the corresponding theoretical curves for the four complex matrix elements.  
Panels \textbf{(b–e)} (1-4  in \textbf{(a)}) correspond to points on the Fermi arc at $(p,q) = (0.1\pi,0.99\pi)$, $(0.5\pi,0.5\pi)$, $(1.0\pi,0.5\pi)$, and $(1.85\pi,0.3\pi)$, respectively.  
Panels \textbf{(f–g)} (5 and 6 in \textbf{(a)}) correspond to off-arc points at $(p,q) = (1.4\pi,0.9\pi)$ and $(0.75\pi,0.8\pi)$, which exhibit a finite imaginary eigenvalue gap.  
These measurements validate the accurate implementation of the non-unitary dynamics and form the basis for the Hamiltonian reconstruction shown in Fig.~\ref{f4}.}
	\label{f3}
\end{figure}

\begin{figure}[htbp]
	\centering
	\includegraphics[width=1.0\linewidth]{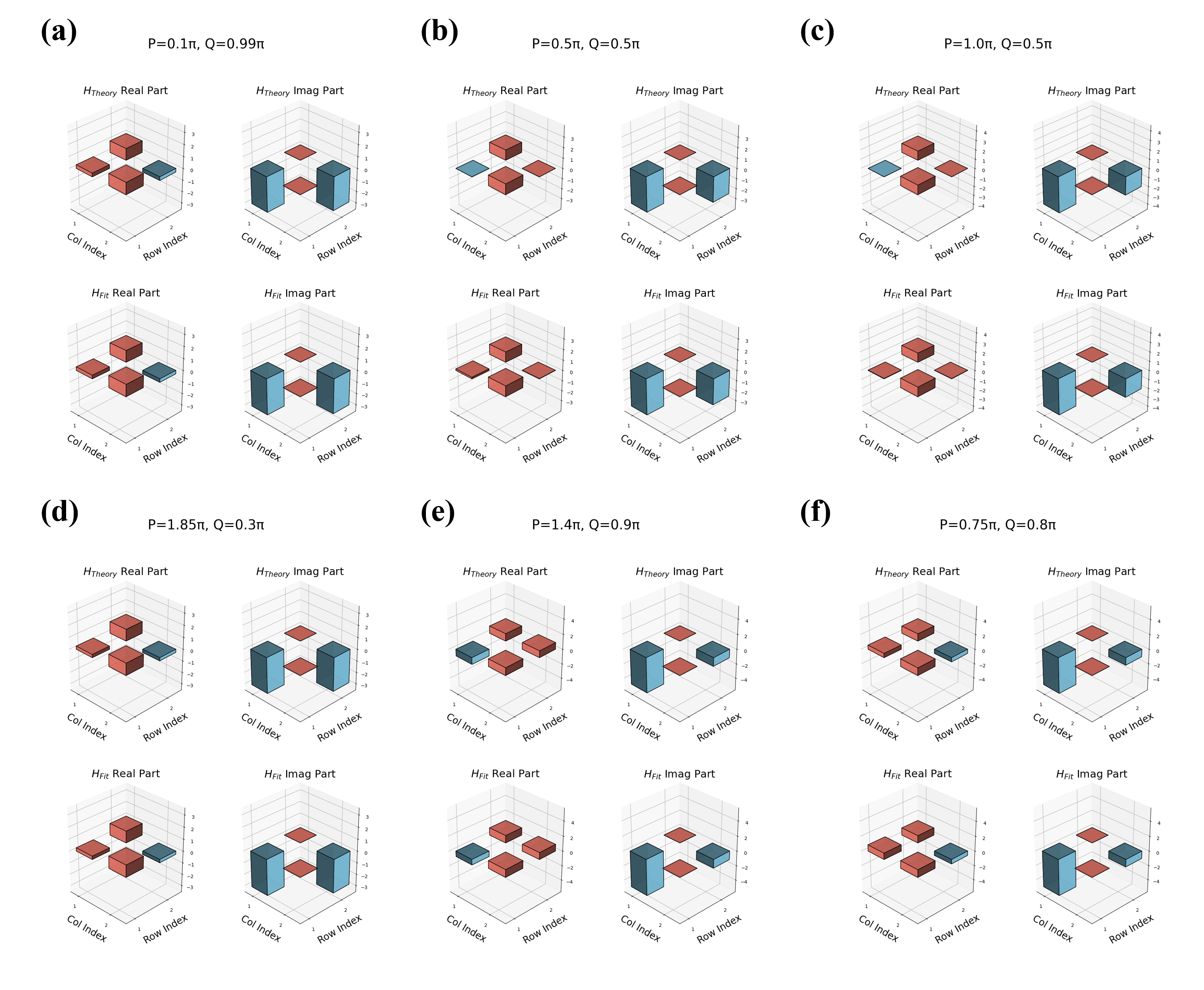}
	\caption{\textbf{Theoretical vs.\ experimentally reconstructed Hamiltonian elements.}
\textbf{(a–f)} Each subpanel compares the real and imaginary parts of the four Hamiltonian matrix elements for six representative parameter points.  
The upper panel in each subfigure shows the theoretically calculated components, while the lower panel shows the corresponding experimentally reconstructed values obtained from fitting the measured time-evolution operators.  
Panels \textbf{(a–d)} correspond to points located on the Fermi arc, whereas panels \textbf{(e–f)} correspond to points in gapped regions away from the arc.}
	\label{f4}
\end{figure}

\Cref{f3}(a) shows the phase diagram for the Hamiltonian in \Cref{eq4} with $c=1$, including the perturbation $e^{0.02 i} H_{\mathrm{EP}}^{\mathrm{\mathbf{K}}^2}$, which sharpens the visibility of the Fermi arc \cite{konig2025exceptional}. The background colormap displays the numerically computed bifurcation of the imaginary parts of the eigenvalues across the $(p,q)$ parameter space with $p\in[0,2\pi]$ and $q\in[0,\pi]$. The Fermi arc appears as the curve where this imaginary eigenvalue gap closes. To probe this structure experimentally, we sampled $11$ points along the arc and $8$ points away from it, covering both gapless and gapped regions of the parameter space. 
Before reconstructing the Hamiltonian, we first validate that the photonic circuit correctly implements the complex time-evolution operator. For each sampled parameter point, we measure the amplitudes and phases of all four elements of $T_{t_n}=e^{-iH' t_n}$ at five evolution times $t_n=0.05\pi, 0.10\pi, 0.15\pi, 0.20\pi,$ and $0.25\pi$. \Cref{f3}(b–g) shows the comparison between the simulated and experimentally measured matrix elements for six representative parameter points. Points~1–4 lie on the Fermi arc, while points~5–6 lie in regions with a large imaginary eigenvalue gap. In all cases, the experimental markers closely follow the predicted evolution, demonstrating faithful reproduction of the non-unitary dynamics on the chip. Having established the accuracy of the measured $T_{t_n}$, we proceed to reconstruct the effective Hamiltonian. For each parameter point, we fit the measured complex matrix elements across the five sampled evolution times to obtain a consistent generator $H'$. \Cref{f4} compares the theoretically predicted real and imaginary parts of the Hamiltonian matrix elements with the experimentally reconstructed values for the same six points. The upper panels show the theoretical components of the Hamiltonian, while the lower panels display the experimentally extracted ones. Across both on-arc and off-arc points, the agreement is excellent, demonstrating that our reconstruction protocol reliably retrieves the underlying non-Hermitian Hamiltonian.
Our measurements provide clear evidence for identifying the Fermi arc by revealing the presence or absence of bifurcation. 

Measuring the unbroken Fermi arc allows us to make statements on the system's phase, and infer the presence of the monopole \cite{konig2025exceptional}.
As there is only one Fermi arc in the system, and it only crosses the orientation-reversing boundary  along the \(p\)-axis [blue path in \Cref{f1}~(a)], we infer that the boundary braids are \(B_p = \sigma_1^{\pm1}\), and \(B_q=1\).
Since the Fermi arc crosses this boundary, or equivalently since the braid along the orientation-reversing boundary is not trivial, there must be exceptional points in the system, whose total charge adds up to \(\pm2\).
This can be inferred from the Fermi arc having a consistent orientation, which can only change at degenerate points and at the orientation-reversing boundary. 
Both ends of the Fermi arc thus are oriented either `out' of, or `into' the fundamental domain.
This non-zero total flow must be sourced by EPs, at which Fermi arcs end, or more fine-tuned monopoles, at which multiple Fermi arcs emerge.

\section*{Discussion}
We have realized a programmable silicon photonic platform for emulating non-Hermitian dynamics using a differential loss scheme, avoiding the need for optical gain. The system implements a two-band non-Hermitian Hamiltonian on a Klein bottle parameter space; a nonorientable manifold that supports exceptional points and Fermi arcs not typically accessible in Hermitian systems. Through precise control of interferometer meshes and ancilla-assisted singular value decomposition, we experimentally reconstructed the time evolution of the system and observed the evolution of Fermi arcs and exceptional points under parameter variation. Our approach leverages thermally tunable Mach-Zehnder interferometers within a CMOS-compatible PIC, allowing high-fidelity sampling of non-unitary transformations. The output intensities from the nominal modes show excellent agreement with theoretical predictions, as quantified by high correlation coefficients across multiple parameter sets. These results suggest that such platforms may serve as a testbed for studying non-Hermitian topology in settings where material gain is challenging to incorporate.

This work contributes to the broader understanding of non-Hermitian phenomena in photonic systems and highlights the potential of reconfigurable linear circuits for emulating Hamiltonians with complex eigenvalue landscapes. The techniques developed here could inform future studies on topological insulators, non-Hermitian skin effects, and nonreciprocal transport in integrated photonics.  Programmable photonic circuits capable of implementing arbitrary linear transformations may provide a pathway toward adaptive optical interconnects, enhanced signal routing flexibility, and improved energy efficiency in large-scale computing systems.

\subsection*{Data Availability}
The data generated in this study is available upon reasonable request from the corresponding author.

\subsection*{Acknowledgments}
JLKK and EJB were supported by the Wallenberg Scholars program (2023.0256), and the project Dynamic Quantum Matter (2019.0068) of the Knut and Alice Wallenberg Foundation, as well as the Göran Gustafsson Foundation for Research in Natural Sciences and Medicine.
ZX, AC, GK and AWE acknowledge support from the Knut and Alice Wallenberg (KAW) Foundation through the Wallenberg Centre for Quantum Technology (WACQT). 
JG acknowledges support from Swedish Research Council (Ref: 2023-06671 and 2023-05288), Vinnova project (Ref: 2024-00466), and the Göran Gustafsson Foundation. 
VZ acknowledges support from the KAW.
AWE acknowledges support from Swedish Research Council (VR) Starting Grant (Ref: 2016-03905). 
\\

\subsection*{Competing Interests}
The authors declare no competing interests.

\subsection*{Author Contributions}
AWE, JG, EJB, and VZ conceived and supervised the project. ZSX, JG, JLKK and AWE designed the experiment. JLKK, EJB and ZSX conducted the theoretical work. ZSX, AC, and RY performed the experiment. ZSX analyzed the experimental data. GK contributed to the calibration of the photonic integrated circuit and the improvement of the experiment setup. ZSX and JLKK wrote the paper, with input from all other authors.

\clearpage

\printbibliography

\clearpage

\clearpage

\clearpage

\clearpage

\clearpage

\end{document}